\documentclass[letterpaper, 10 pt, conference]{ieeeconf}  

\IEEEoverridecommandlockouts                              
                                                          
\DeclareUnicodeCharacter{200B}{{\hskip 0pt}}
\usepackage{graphics} 
\usepackage{epsfig} 
\usepackage{mathptmx} 
\usepackage{times} 
\usepackage{amsmath} 
\usepackage{amssymb}  
\usepackage{caption}
\usepackage{subcaption}
\usepackage{float}
\usepackage{url}
\usepackage{multirow}
\usepackage[export]{adjustbox}
\usepackage{footnote}

\makeatletter
\let\NAT@parse\undefined
\makeatother
\usepackage{hyperref}
\hypersetup{
    colorlinks=true,
    linkcolor=blue,
    filecolor=magenta,      
    urlcolor=cyan,
    pdftitle={Overleaf Example},
    pdfpagemode=FullScreen,
    }

\title{\LARGE \bf
A novel multi-view deep learning approach for BI-RADS and density assessment of mammograms
}

\author{Huyen T. X. Nguyen$^{1,\dag}$, Sam B. Tran$^{1,\dag}$, Dung B. Nguyen$^{1}$, Hieu H. Pham$^{2,3,*}$, Ha Q. Nguyen$^{1,2}$%
\thanks{$^{\dag}$ The ﬁrst two authors contributed equally to this work.}%
\thanks{$^{*}$ Corresponding author: \texttt{hieu.ph@vinuni.edu.vn} (Hieu Pham).}%
\thanks{$^{1}$ Smart Health Center, VinBigdata, Hanoi, Vietnam}%
\thanks{$^{2}$ College of Engineering \& Computer Science, VinUniversity, Hanoi, Vietnam}%
\thanks{$^{3}$ VinUni-Illinois Smart Health Center, VinUniversity, Hanoi, Vietnam}%
}

\begin{document}

\maketitle
\thispagestyle{plain}
\pagestyle{plain}

\begin{abstract}
Advanced deep learning (DL) algorithms may predict the patient’s risk of developing breast cancer based on the Breast Imaging Reporting and Data System (BI-RADS) and density standards. Recent studies have suggested that the combination of multi-view analysis improved the overall breast exam classification. In this paper, we propose a novel multi-view DL approach for BI-RADS and density assessment of mammograms. The proposed approach first deploys deep convolutional networks for feature extraction on each view separately. The extracted features are then stacked and fed into a Light Gradient Boosting Machine (LightGBM) classifier to predict BI-RADS and density scores. We conduct extensive experiments on both the internal mammography dataset and the public dataset Digital Database for Screening Mammography (DDSM). The experimental results demonstrate that the proposed approach outperforms the single-view classification approach on two benchmark datasets by huge \textit{F1}-score margins (+5\% on the internal dataset and +10\% on the DDSM dataset). These results highlight the vital role of combining multi-view information to improve the performance of breast cancer risk prediction.

{\textbf{\textit{Index Terms}}}\textemdash Mammogram, multi-view deep learning, BI-RADS and density classification.
\end{abstract}
\section{INTRODUCTION}
\thispagestyle{empty}
Breast cancer has now beat lung cancer to become the most commonly diagnosed cancer, according to statistics released by the International Agency for Research on Cancer (IARC) in December 2020~\cite{iarc}. It is estimated that there were 2.3 million women diagnosed with breast cancer and 685,000 deaths globally in 2020~\cite{who}. Symptoms frequently appear at the later stage of breast cancer, but treatments could be confronted with challenges in this period. Hence, regular breast cancer screening plays a crucial role in the early detection of breast tumors. Mammography, a type of X-ray examination for breasts, is utilized in computer-aided diagnosis (CADx) systems to enhance radiologists' efficiency. A typical mammogram consists of four views: R-CC (right craniocaudal), L-CC (left craniocaudal), R-MLO (right mediolateral oblique), and L-MLO (left mediolateral oblique). In clinical practice, physicians usually use these views for evaluating breast cancer risk. In particular, BI-RADS score is used as a risk assessment and quality assurance tool that supplies a widely accepted lexicon and reporting schema for imaging of the breast~\cite{birad}. This standard contains seven assessment categories: BI-RADS 0 (incomplete), BI-RADS 1 (negative), BI-RADS 2 (benign), BI-RADS 3 (probably benign), BI-RADS 4 (suspicious for malignancy), BI-RADS 5 (highly suggestive of malignancy), and BI-RADS 6 (known biopsy-proven malignancy). Besides, breast density refers to the amount of fibroglandular tissue in a breast relative to fat; its four descriptors include A (almost entirely fatty breast), B (scattered areas of fibroglandular density), C (heterogeneously dense breast), and D (extremely dense breast)~\cite{density}.

Previously, many machine learning approaches have been suggested to classify and detect breast cancer using single-view information~\cite{6399758,8964871}, based on texture descriptors or DL networks. Recently, several studies showed that multi-view approaches~\cite{8861376,8897609,Geras2017HighResolutionBC} improved the diagnosis of breast cancer. The primary technique behind these approaches is building an end-to-end DL model to classify mammograms' pathology. This strategy first extracts features from each view independently and then combines four screening mammography views to produce predictions. In addition, we observed that most current work in breast cancer diagnosis focuses on the discrimination of a mammogram exam as malignant or benign. Nevertheless, few methods could classify mammograms into multi-output. This motivates us to build a DL system that is able to provide a multi-label output, including 5 BI-RADS categories (BI-RADS 1-5) and 4 density classes (A-D). Our experiments show that the proposed multi-view approach consistently outperforms the single-view approach on internal and external benchmark datasets.

Our work makes two main contributions. \textbf{First}, we develop a multi-label DL system that can automatically classify the BI-RADS density of mammograms. To our knowledge, we show for the first time in this work that a DL model trained on a large-scale, annotated dataset can predict both BI-RADS and density scores accurately at the same time. \textbf{Second}, the proposed approach is based on a novel two-stage multi-view classifier. The first stage learns feature extractors for each view individually. At the second stage, learned features are fused and afterward trained by a LightGBM classifier~\cite{LightGBM} to predict BI-RADS and density scores. Experimental results show that our multi-view approach outperforms the single-view model up to 5$\%$ in terms of \textit{F1}-score. External evaluation on the DDSM dataset~\cite{DDSM} also demonstrates the effectiveness of the proposed approach with an increase of 10$\%$ in pathology classification compared to the single view approach. The rest of the paper is organized as follows. The proposed approach is presented in Section~\ref{sec:model}. The experiments are provided in Section~\ref{sec:ExperimentResult}. Finally, Section~\ref{sec:conclusion} discusses the experimental results and concludes the work as well as presents its perspectives.

\section{PROPOSED APPROACH}
\thispagestyle{empty}
\label{sec:model}
\subsection{Deep Learning-based BI-RADS Density Classification}
\label{ssec:subhead42}

The proposed BI-RADS density classification architecture has two stages, as illustrated in Figure~\ref{fig:2c}.  The first stage used CNNs to learn latent features from mammograms. The second stage then applied a LightGBM as a classifier to predict BI-RADS and density scores. To verify the effectiveness of the proposed multi-view model, we compare it with a single view approach (Figure~\ref{fig:2a}). 
 
\subsection{Image Preprocessing}
\label{ssec:subhead41}
Most mammogram scans have a black background without any information, which is hugely resource-consuming for training. Hence, we built an automatic detector using YOLOv5~\cite{yolov5} to accurately localize the relevant region of the breast from the original scans. To train the detector, we manually annotated 2,000 mammograms, from which 1,600 scans were used for training and 400 scans for validation. The trained model reported an mAP of 0.995 on the validation set and could correctly predict the breast's bounding boxes in unseen mammograms.
 
\subsection{Multi-view Feature Extraction}
\label{sssec:subsubhead421}
According to the feature extraction process, the supervised deep CNNs have been implemented and then removed the fully connected layer at the end of the architecture to get the hidden representations from the input images. The hidden representation is a tensor with dimensions $\textit{H} \times \textit{W} \times \textit{C}$ that is down-sampled from the original dimensions. Specifically, we used ResNet-34~\cite{he2016deep} and EfficientNet-B2~\cite{tan2019efficientnet} for the feature extraction step. To train ResNet-34 extractor, all input images were resized to $\textit{H} \times \textit{W} = \textit{512} \times \textit{512}$. Additionally, each mammogram has been duplicated from the original grayscale channel to three channels.  After passing the trained extractor, we obtained the feature tensor with dimensions of $\textit{H} \times \textit{W} \times \textit{C} = \textit{32} \times \textit{24} \times \textit{512}$. Finally, this hidden representation was averaged-pooled over the spatial dimensions to get a 512-dimensional vector.  Likewise,  to train EfficientNet-B2 model, the input was $ \textit{H} \times \textit{W} \times \textit{C} = \textit{1024} \times \textit{768} \times \textit{3}$. The hidden tensor was  $\textit{H} \times  \textit{W} \times \textit{C} = \textit{32} \times \textit{24}  \times \textit{1408}$ and the hidden vector was in $\textit{1408}$-dimensional format.

\begin{figure*}
     \centering
     \begin{subfigure}[b]{0.45\textwidth}
         \centering
         \includegraphics[scale=.24]{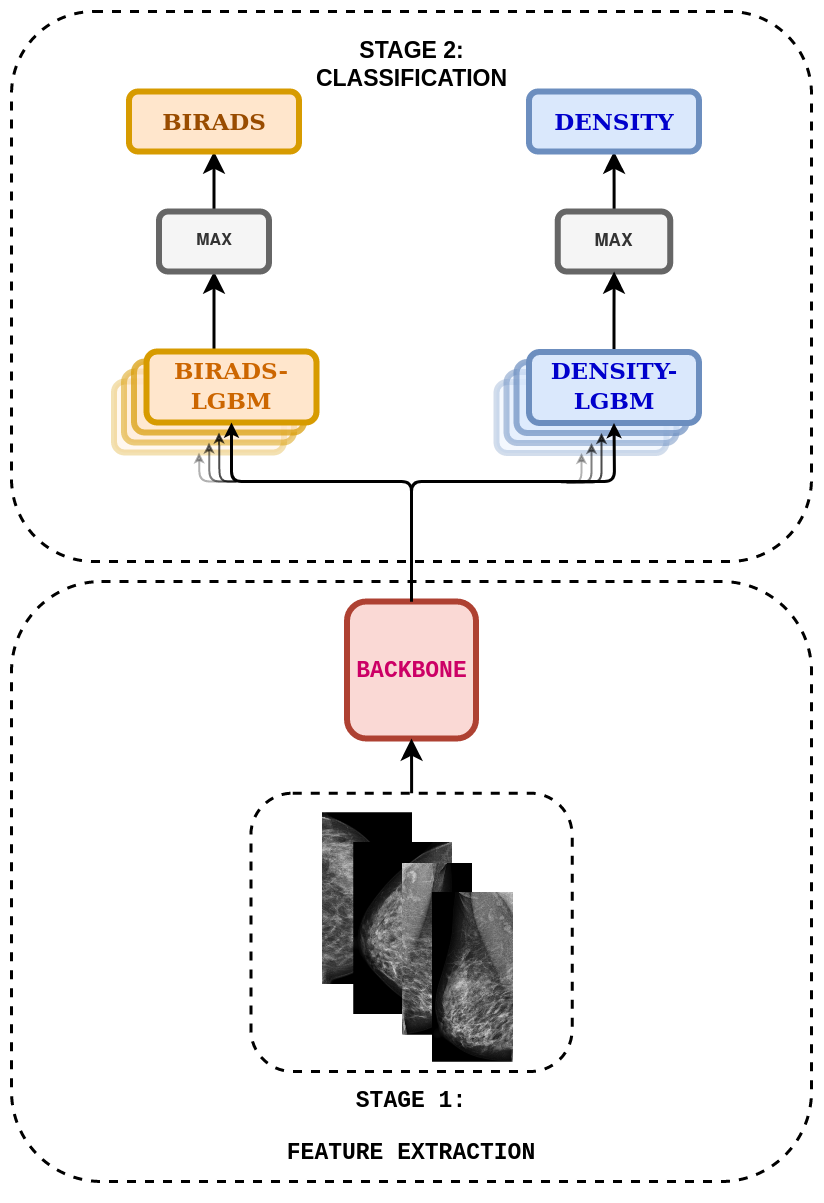}
         \caption{Single-view model}
         \label{fig:2a}
     \end{subfigure}
     \begin{subfigure}[b]{0.45\textwidth}
         \centering
         \includegraphics[scale=.24]{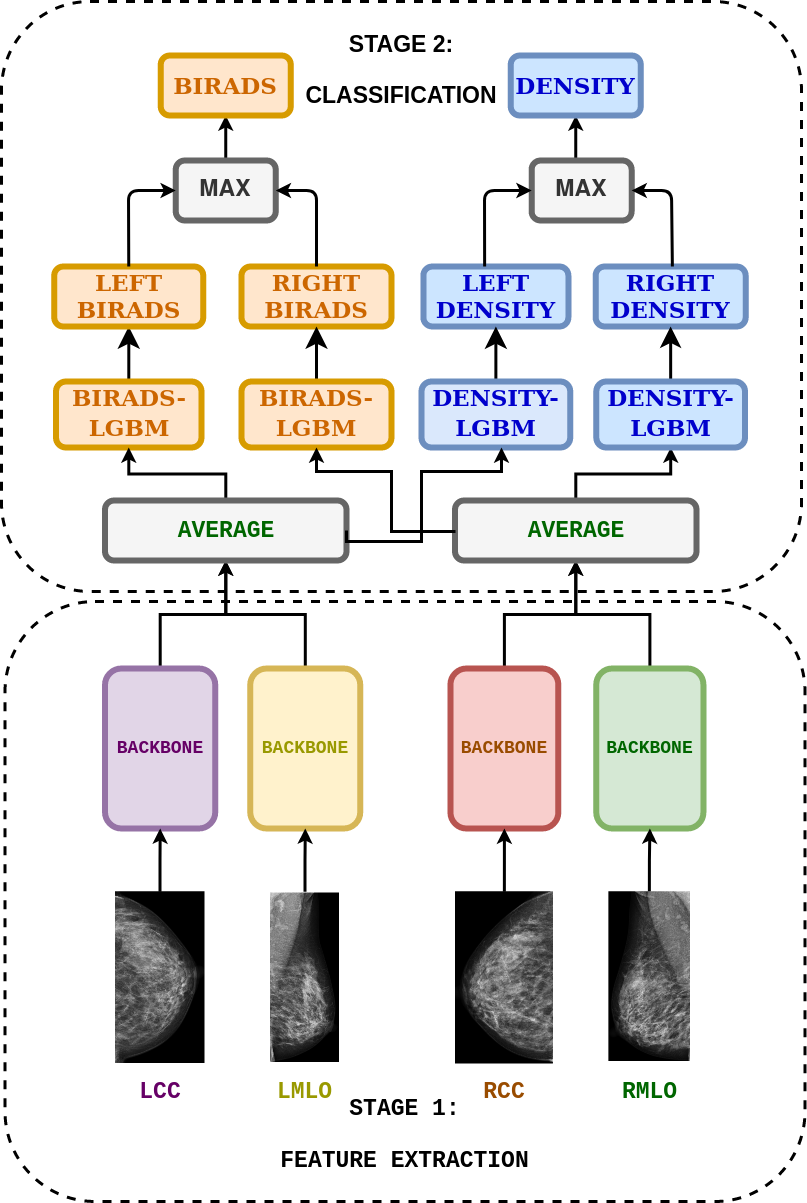}
         \caption{Multi-view model (\textbf{ours})}
         \label{fig:2c}
     \end{subfigure}
        \caption{\small{Illustration of the single view approach (a) and our multi-view approach (b) for BI-RADS and density assessment of mammograms using deep neural networks. Unlike a traditional single-view approach that takes single breast views as inputs during training, the proposed multi-view model takes four breast views as inputs and learns features independently. After extracting hidden features from each view separately, CC and MLO features were combined and fed into a LightGBM classifier to predict the outcomes.}}
        \label{fig:models}
\end{figure*}

\begin{table*}[ht!]
\centering
\scriptsize{
    \caption{Description of the private dataset used for model development and validation.}
    \label{tab:1}
    \begin{tabular}{||c||c|c|c|c|c||c|c|c|c|c||c|c|c|c|c||}
    \hline
    \textbf{Data}    & \multicolumn{5}{c||}{\textbf{Training set}}       & \multicolumn{5}{c||}{\textbf{Validation set}}    & \multicolumn{5}{c||}{\textbf{Test set}}       \\
    \hline
    \textbf{BI-RADS} & \textbf{LCC}   & \textbf{RCC}   & \textbf{LMLO}  & \textbf{RMLO}  & \textbf{Total}  & \textbf{LCC}   & \textbf{RCC}   & \textbf{LMLO}  & \textbf{RMLO}  & \textbf{Total} & \textbf{LCC}   & \textbf{RCC}   & \textbf{LMLO}  & \textbf{RMLO}  & \textbf{Total} \\
    \hline
    \hline
    \textbf{1}       & 4,001 & 4,036 & 4,006 & 4,038 & 16,081 & 846   & 871   & 848   & 871   & 3,436 & 858   & 853   & 859   & 854   & 3,424 \\
    \hline
    \textbf{2}       & 1,320 & 1,337 & 1,437 & 1,467 & 5,561  & 289   & 280   & 287   & 280   & 1,136 & 288   & 281   & 286   & 280   & 1,135 \\
    \hline
    \textbf{3}       & 428   & 416   & 463   & 449   & 1,756  & 87    & 76    & 87    & 76    & 326   & 87    & 92    & 87    & 91    & 357   \\
    \hline
    \textbf{4}       & 271   & 237   & 729   & 663   & 1,900  & 34    & 35    & 34    & 36    & 139   & 31    & 37    & 32    & 38    & 138   \\
    \hline
    \textbf{5}       & 104   & 86    & 279   & 223   & 692    & 10    & 4     & 10    & 3     & 27    & 7     & 8     & 7     & 8     & 30    \\
    \hline
    \textbf{Total}   & 6,124 & 6,112 & 6,914 & 6,840 & 25,990 & 1,266 & 1,266 & 1,266 & 1,266 & 5,064 & 1,271 & 1,271 & 1,271 & 1,271 & 5,064 \\
    \hline
    \end{tabular}}

\end{table*}

\subsection{BI-RADS \& Density Classification}
\label{sssec:subsubhead422} 
The input of this phase is the feature vectors from the feature extraction step. We trained a LightGBM classifier on $\textit{N} \times \textit{512}$-dimensional vectors provided by ResNet-34 and $\textit{N} \times \textit{1408}$-dimensional vectors provided by EfficientNet-B2, where $\textit{N}$ is the number of training images. The LightGBM is a gradient boosting framework based on tree algorithms implemented to deliver results faster, reduce memory usage, and improve accuracy. The main idea behind multi-view (Figure~\ref{fig:2c}) architecture was to use a fusion of multiple hidden information from four views to train an efficient classifier.  The four backbones in the feature extraction step are trained for each view of the dataset independently. At this combination technique, we averaged the  L-CC  and L-MLO  feature vectors, and the R-CC and R-MLO feature vectors.  Then all the average vectors were used as inputs for LightGBM to provide the confident scores of BI-RADS and density.

\section{EXPERIMENTS \& RESULTS}
\label{sec:ExperimentResult}
\thispagestyle{empty}
\subsection{Dataset Preparation}
\label{sec:data}
We evaluate the proposed method on a part of VinDr-Mammo dataset~\cite{nguyen2022vindr} and DDSM dataset. The private dataset was retrospectively collected from Hanoi Medical University Hospital from 2018 to 2020. Each image was assigned to a team of three radiologists specializing in breast imaging for multilabel annotation (BI-RADS 1-5 and breast density A-D). In total, the dataset includes 36,138 screening mammogram images from 9,911 studies. There are 8509 four-view studies, which consist of four-view images (L-CC, L-MLO, R-CC, R-MLO). These exams were divided into three groups by the multilabel stratification method~\cite{stratify}: training set (5,792 studies), validation set (1,266 studies), and test set (1,271 studies). Descriptions of three sets on the private dataset are provided in Table~\ref{tab:1}. For the DDSM dataset, the number of exams that are multilabel-stratified in training, validation, test set is 1,822; 391; and 391, respectively. Each study involves two-view images of each breast, along with pathology label and density information. There are four types of labels, \textit{normal} - \textit{benign} - \textit{cancer} - \textit{benign with callback} - \textit{benign without callback}. The DDSM training set contains 486 normal, 704 benign, and 632 cancer cases. The figures of normal/benign/cancer studies in the validation set and test set are 105/147/139 and 104/145/142, respectively. For both datasets, each breast's pathology/BI-RADS label is the maximum label of its CC and MLO images, while the density of these two view images certainly has the same information. 
\subsection{Training \& Evaluation Metrics }
\label{ssec:subhead51}
This study was built on PyTorch version 1.8.1 (\url{https://pytorch.org/}), and used a machine with an Nvidia GTX 1080 GPU. We trained the feature extractors using SGD optimizer~\cite{ruder2016overview} with momentum = 0.9 and cosine annealing learning rate~\cite{CALR}. The cross-entropy function was used to calculate the error. For model evaluation, we used \textit{F1}-score on the 5-class BI-RADS level and 4-class density level. \textit{F1}-score is the harmonic mean of precision and recall. For multi-class problems, macro-\textit{F1} score, which is defined as the mean of class-wise \textit{F1}-scores could be used. As each study's left side and right side might have characteristic differences, the results are appraised for left breasts, right breasts, and study-level for breast diagnosis (BI-RADS types or pathology labels). The evaluation of density cases is based on the left, right, and side-level.

For a fair comparison, we used the same network architecture (ResNet-34/EfficientNet-B2) as the feature extractor with a fixed-size input image for both the single-view and multi-view models. The number of epochs was set to 50, and the training process stopped in case there was no improvement in \textit{F1}-score of the validation set after 15 consecutive epochs by an early stopping callback. The model which acquired the best \textit{F1}-score for validation would be selected as the best feature extractor for the LightGBM classifier. Evaluation metrics of these different network architectures are assessed on the test set.

\begin{table*}[ht!]
\centering 
\small{
    \caption{Quantitative results using \textit{F1}-score of different architectures on the private test dataset.}
    \label{tab:2}
    \begin{tabular}{||c||c||c|c|c||c|c|c||c|c|c||}
    \hline
    \textbf{Backbone} & \textbf{Model} & \multicolumn{3}{c||}{\textbf{Single-view model}} & %
    \multicolumn{3}{c||}{\textbf{Multi-view model}}\\
    \hline
    \hline
    \multirow{13}{*}{\textbf{ResNet-34}} & \textbf{BI-RADS} & \textbf{Left} & \textbf{Right} & \textbf{Study} & \textbf{Left} & \textbf{Right} & \textbf{Study} \\ 
    \cline{2-8} & 1 & 0.78 & 0.78 & 0.65 & 0.85 & 0.85 & 0.78 \\ 
    \cline{2-8} & 2 & 0.52 & 0.51 & 0.53 & 0.53 & 0.51 & 0.56 \\ 
    \cline{2-8} & 3 & 0.17 & 0.21 & 0.23 & 0.16 & 0.29 & 0.27 \\ 
    \cline{2-8} & 4 & 0.13 & 0.18 & 0.18 & 0.14 & 0.25 & 0.20 \\ 
    \cline{2-8} & 5 & 0.62 & 0.67 & 0.64 & 0.73 & 0.71 & 0.72 \\ 
    \cline{2-8} & Macro-\textit{F1} & 0.4456 & 0.4688 & \textcolor{red}{\textbf{0.4473}} & 0.4813 & 0.525 & \textcolor{red}{\textbf{0.5063}} \\ \cline{2-8} 
    & \textbf{Density} & Left & Right & Side & Left & Right & Side \\ 
    \cline{2-8} & A & 0.27 & 0.12 & 0.19 & 0.40 & 0.33 & 0.37 \\ 
    \cline{2-8} & B & 0.48 & 0.54 & 0.51 & 0.54 & 0.62 & 0.58 \\ 
    \cline{2-8} & C & 0.70 & 0.71 & 0.71 & 0.73 & 0.74 & 0.74 \\ 
    \cline{2-8} & D & 0.59 & 0.61 & 0.60 & 0.59 & 0.60 & 0.60 \\ 
    \cline{2-8} & Macro-\textit{F1} & 0.5118 & 0.4973 & \textcolor{red}{\textbf{0.504}} & 0.5666 & 0.5722 & \textcolor{red}{\textbf{0.5698}} \\ 
    \hline
    \hline
    \multirow{13}{*}{\textbf{EfficientNet-B2}} & \textbf{BI-RADS} & \textbf{Left} & \textbf{Right} & \textbf{Study} & \textbf{Left} & \textbf{Right} & \textbf{Study} \\ 
    \cline{2-8} & 1 & 0.85 & 0.88​& 0.80 & 0.88​& 0.89​& 0.82 \\ 
    \cline{2-8} & 2 & 0.59​& 0.63​& 0.63 & 0.63​& 0.62​& 0.66​\\ 
    \cline{2-8} & 3 & 0.30 & 0.29 & 0.35 & 0.37 & 0.35 & 0.43 \\ 
    \cline{2-8} & 4 & 0.30 & 0.28 & 0.28 & 0.19 & 0.29 & 0.28 \\ 
    \cline{2-8} & 5 & 0.77 & 0.67 & 0.72 & 0.83 & 0.55 & 0.70 \\ 
    \cline{2-8} & Macro-\textit{F1} & 0.5617 & 0.5502 & \textcolor{red}{\textbf{0.5577}} & 0.5802 & 0.5393 &   \textcolor{red}{\textbf{0.5759}} \\ \cline{2-8} 
    & \textbf{Density} & \textbf{Left} & \textbf{Right} & \textbf{Side} & \textbf{Left} & \textbf{Right} & \textbf{Side} \\ 
    \cline{2-8} & A & 0.25 & 0.36 & 0.32 & 0.53 & 0.48 & 0.50 \\ 
    \cline{2-8} & B & 0.57 & 0.53 & 0.55 & 0.62 & 0.63 & 0.62 \\ 
    \cline{2-8} & C & 0.72 & 0.74 & 0.73 & 0.76 & 0.78 & 0.77 \\ 
    \cline{2-8} & D & 0.61 & 0.62 & 0.61 & 0.56 & 0.58 & 0.57 \\ 
    \cline{2-8} & Macro-\textit{F1} & 0.5386 & 0.5615 & \textcolor{red}{\textbf{0.5525}} & 0.6161 & 0.6184 & \textcolor{red}{\textbf{0.6165}} \\ 
    \hline
    \end{tabular}}
    
\end{table*}

\begin{table*}[ht!]
    \centering
    \small{
    \caption{Quantitative results using \textit{F1}-score of EfficientNet-B2 on the DDSM dataset.}
    \label{tab:3}
    \begin{tabular}{||c||c||c|c|c||c|c|c||c|c|c||}
    \hline
    \textbf{Backbone} & \textbf{Model} & \multicolumn{3}{c||}{\textbf{Single-view model}} & %
    \multicolumn{3}{c||}{\textbf{Multi-view model}}\\
    \hline
    \hline
    \multirow{8}{*}{\textbf{EfficientNet-B2}} & \textbf{Label} & \textbf{Left} & \textbf{Right} & \textbf{Study} & \textbf{Left} & \textbf{Right} & \textbf{Study} \\ 
    \cline{2-8} & Normal & 0.74 & 0.70 & 0.67 & 0.84 & 0.77 & 0.78 \\ 
    \cline{2-8} & Benign & 0.54 & 0.56 & 0.44 & 0.65 & 0.65 & 0.62 \\ 
    \cline{2-8} & Malignant & 0.58 & 0.59 & 0.61 & 0.55 & 0.49 & 0.61 \\ 
    \cline{2-8} & Macro-\textit{F1} & 0.6186 & 0.6173 & \textcolor{red}{\textbf{0.5733}} & 0.6806 & 0.6353 & \textcolor{red}{\textbf{0.6702}} \\
    \cline{2-8}& \textbf{Density} & \textbf{Left} & \textbf{Right} & \textbf{Side} & \textbf{Left} & \textbf{Right} & \textbf{Side} \\
    \cline{2-8} & A & 0.08 & 0.13 & 0.11 & 0.07 & 0.18 & 0.13 \\ 
    \cline{2-8} & B & 0.47 & 0.41 & 0.44 & 0.52 & 0.42 & 0.48 \\ 
    \cline{2-8} & C & 0.31 & 0.29 & 0.30 & 0.38 & 0.39 & 0.39 \\ 
    \cline{2-8} & D & 0.46 & 0.43 & 0.45 & 0.42 & 0.42 & 0.42 \\ 
    \cline{2-8} & Macro-\textit{F1} & 0.3319 & 0.3145 & \textcolor{red}{\textbf{0.3233}} & 0.3472 & 0.3550 & \textcolor{red}{\textbf{0.3544}} \\ 
    \hline
    \end{tabular}}
\end{table*}

\subsection{Experimental Results}
\thispagestyle{empty}
\label{ssec:subhead52}
Table~\ref{tab:2} illustrates the experiment results of two models with different feature extractors (ResNet-34/ EfficientNet-B2) on the private dataset. We observed that the performance of the proposed multi-view model had surpassed the single view model for both BI-RADS and density classification. In the case of BI-RADS classifiers with ResNet-34 backbone, all BI-RADS classes of multi-view model achieve the best \textit{F1}-score compared to single-view. For multi-view architecture, its metric is approximately 6$\%$ higher than the single-view model. Similarly, with EfficientNet-B2, the multi-view model impacts more positively to BI-RADS prediction than the single-view model with a \textit{F1}-score of 57.59$\%$. In the case of density categorization, evaluation on the test set of single-view is inferior to the multi-view model. Multi-view results of feature extractor ResNet-34 and EfficientNet-B2, which were calculated on exams, are 56.98$\%$ and 61.65$\%$ respectively. According to a vast number of executed experiments, the effectiveness of multi-view architecture on our private dataset is proved. Furthermore, we also present the respective performance for two different model architectures on the DDSM dataset. The most important purpose of this experiment is to show the efficiency of the multi-view model on another data distribution. We use EfficientNet-B2 as a backbone to extract the feature representation of DDSM samples. According to our result listed in Table~\ref{tab:3}, our proposed multi-view architecture achieves a higher \textit{F1}-score than single-view by approximately 9.69$\%$ at study-level in the classification of pathology and 3.11$\%$ at side-level in density classification. The combination of four-view information in mammography can surpass the performance of several variant models of just one view as demonstrated by the results in Table~\ref{tab:2},~\ref{tab:3}. In conclusion, this multi-view approach significantly impacts BI-RADS and density classification. Description in Section~\ref{sec:data} shows that the pathology balance of DDSM dataset might be better than the private dataset. As a result, the improvement score between multi-view model and single-view model on DDSM dataset is superior on pathology level. 

\section{DISCUSSION AND CONCLUSION}
\thispagestyle{empty}
\label{sec:conclusion} 
We introduced in this paper a novel multi-view strategy to classify breast density and estimate BI-RADS scores from mammogram exams. We demonstrated empirically on both the private and public DDSM datasets that the combined information from MLO and CC views improves diagnostic accuracy compared to a single view approach. This opens a new promising approach for building and developing an effective model in early breast cancer detection. We are currently expanding this study by investigating new feature combination strategies to generate more discriminative features for BI-RADS and density classification tasks.

\section*{COMPLIANCE WITH ETHICAL STANDARDS}
Our work follows all applicable ethical research standards and laws. The study has been reviewed and approved by the institutional review
board (IRB) of the hospital. The need for obtaining informed patient consent was waived because this work did not impact clinical care.

\section*{ACKNOWLEDGMENT}
This study was supported by Smart Health Center, VinBigdata. We would like to acknowledge Hanoi Medical University Hospital for providing us access to their image databases. In particular, we thank all of our radiologists who participated in this project.


\end{document}